# Privacy-Preserving Blockchain Based Federated Learning with Differential Data Sharing

Anudit Nagar

*Abstract* — For the modern world where data is becoming one of the most valuable assets, robust data privacy policies rooted in the fundamental infrastructure of networks and applications are becoming an even bigger necessity to secure sensitive user data. In due course with the ever-evolving nature of newer statistical techniques infringing user privacy, machine learning models with algorithms built with respect for user privacy can offer a dynamically adaptive solution to preserve user privacy against the exponentially increasing multi-dimensional relationships that datasets create. Using these privacy aware ML Models at the core of a Federated Learning Ecosystem can enable the entire network to learn from data in a decentralized manner. By harnessing the ever-increasing computational power of mobile devices, increasing network reliability and IoT devices revolutionizing the smart devices industry, and combining it with a secure and scalable, global learning session backed by a blockchain network with the ability to ensure on-device privacy, we allow any Internet enabled device to participate and contribute data to a global privacy-preserving, data sharing network with blockchain technology even allowing the network to reward quality work. This network architecture can also be built on top of existing blockchain networks like Ethereum and Hyperledger, this lets even small startups build enterprise ready decentralized solutions allowing anyone to learn from data across different departments of a company, all the way to thousands of devices participating in a global synchronized learning network.

*Keywords* — *Blockchain Technology, Differential Privacy, Ethereum, Encryption, Federated Learning, Incentivized Learning, IoT, Scalable, Secure.*

## I. Understanding Data Privacy

### A. Segmenting and Understanding Data

Databases of user information, user behaviors and inferences made from these data points by various models, can be primarily categorized from a privacy standpoint into the following categories [1] :

- Personally Identifiable Information (PII): Data points that directly correlate to users. For example - Unique Identification Numbers, Driving License Numbers.
- Quasi-Identifiers (QI): These are data points which are not necessarily on their own vital, but when paired with other QIs can prove to be vital in identification. For example – Pin Code, Gender, Birthday.
- Sensitive Columns (SC): These are not PII or QI but still need to be protected as they are critical details about an individual. For example – Medical Diagnosis, Salary.
- Non-sensitive Columns: These are data points that do not fit the criteria for PII, QI or SC. For example – Country.

### B. Security Challanges

Storing and releasing entire datasets is one the biggest problems that data scientists are facing. Rigorous methods of privacy are proving computationally futile for working with high dimensional data. Starting simple, the way of preventing private data of users being released in datasets would be to remove the data points that directly link to the user (PIIs) called data anonymization [2] and breaking the entire dataset into smaller subsets creating micro-data which is characterized by high dimensionality and sparsity [3].

Many models of privacy have since arose for maintaining data privacy, most approaches building upon the core idea of k-anonymity [1]. In a such a model, the practice is to remove the personally identifiable information (PII) and generalize the quasi-identifiers (QI) until each generalized record is indistinguishable from at least *k-1* other generalized records based on any sensitive identifier.

Such a method of data sanitization also known as masking does not guarantee privacy, as in recent years de-anonymization attacks have proved to be disastrous to such a technique of user privacy wherein an adversary has little background knowledge about certain records and is able to filter these records out by correlating data from other sources de-anonymizing the records with the help of Quasi Identifiers. An example of such an attack would be the de-anonymization of the Netflix Prize Dataset [3] where researchers were able to de-anonymize records of the Netflix Dataset by correlating entries from Internet Movie Database (IMDb). Thus, while k-anonymity offers protection against 'membership inference' attacks, it does not protect against 'attribute inference' attacks emerging from homogeneity of equivalence groups.

'L-Diversity' [4] aims to solve this by making sure that all equivalence groups have 'attribute diversity'. That is, it ensures that subsets of the dataset that have the same value for a QI have sufficient diversity of the sensitive attribute (at least '*l*' different values). This must be maintained across all equivalence groups working in conjunction with k-anonymity. This is still a vulnerable approach as attackers can exploit the relationships between the attribute values which have very different levels of sensitivity to extract private information from the dataset.

'T-Closeness' [5] tries to solve these issues by actively keeping the distribution of each sensitive attribute in an equivalence group 'close' to its distribution in the original dataset. 't-closeness' is the distance between the distributions of a sensitive attribute and the attribute in the entire dataset in an equivalence group which is no more than the threshold 't'. It utilizes the notion of Earth Mover's Distance (EMD) to represent distance between distributions annulling the effects of attribute sensitivities.

This is where another layer of privacy protection comes into play termed as plausible deniability [6]. Plausible deniability states that an output record can be released only if certain amount of input records are indistinguishable, up to a specified privacy parameter. Such a technique helps in sharing of data in an anonymized fashion, which theoretically to a certain extent cannot be traced back. To maintain plausible deniability in a system for a privacy parameter *k > 0*, there need to be at least *k* input records that produce the output record with similar probability.

## C. Utilizing Differential Privacy

Differential privacy [7] is the method by which a controlled amount of noise is introduced during processing of data while making sure the inferences derived from that data are accurate enough to be utilized. It creates a mathematical framework used to analyze the extent to which any general machine learning model remembers PIIs which it used to derive inferences from a particular dataset, hence giving us the ability to grade models based on privacy and their ability to preserve it making sure that these datapoints do not trace back to a certain individual revealing sensitive information about the individual.

The Laplacian mechanism [7] is a $\varepsilon$-differentially private mechanism for queries $\mathcal{F}$ with answers $\mathcal{F}(\mathcal{D}) \in \mathbb{R}^p$, in which *sensitivity* [8] (Definition 1) plays an important role.

Definition 1: Given a query $\mathcal{F}$ and a normal function $\|.\|$ over the range of $\mathcal{F}$, the *sensitivity* $s(\mathcal{F}, \|.\|)$ is defined as

$$s(\mathcal{F}, \|.\|) = \max_{d(\mathcal{D},\mathcal{D}')=1} \|\mathcal{F}(\mathcal{D}) - \mathcal{F}(\mathcal{D}')\|$$

The normal function $\|.\|$ is either $L_1$ or $L_2$ norm.

The Laplacian mechanism given a query $\mathcal{F}$ and a normal function over the range of $\mathcal{F}$, the random function $\tilde{\mathcal{F}}(\mathcal{D}) = \mathcal{F}(\mathcal{D}) + \eta$ satisfies $\varepsilon$ - differential privacy. Here $\eta$ is a random variable whose probability density function is $p(\eta) \propto e^{-\frac{\varepsilon\|\eta\|}{s(\mathcal{F},\|.\|)}}$.

One of the real-world implementations of this has been documented by Apple [9] focusing on estimating frequencies of elements, for example – emojis. This approach is broken down into a few steps; at first the information is privatized using local differential privacy to assure that the servers do not receive any identifier that could relate back to the user. Next, the information is then sent, via secure channels to the servers where IP identifiers and other metadata is removed to ensure privacy. Third, the data is taken and aggregated after which the data is restrictively shared. These data points are opt-in and are received as a subset of the original data on the device, at specified intervals.

The primary distinctions that differential privacy offers from colloquial notions of privacy is a procedure to quantify the privacy risk incurred when a differentially private system is set in place termed as the privacy budget or privacy loss and is denoted by $\varepsilon$ (epsilon). This enables us to measure the risk an individual's data is put at when their data is included in the statistical study. The higher the value of $\varepsilon$, the less careful the algorithm is towards the protection of an individual's data privacy. This approach however fails to address certain issues [10].

The privacy budget operates on the principle of differential privacy loss which increases drastically when the data submission of an individual user increases more than once as the overall loss in privacy is the summation of individual losses happening over each submission, and with the privacy budget and $\varepsilon$ (epsilon) values being kept a secret, this leads to a compounded effect jeopardizing an individual's privacy.

Machine Learning research and Differential Privacy research are beginning to find common ground where in privacy is becoming an important part of ML research. This interconnect is a family of algorithms called Private Aggregation of Teacher Ensembles (PATE) [11]. The PATE framework carefully coordinates learning across several ML models to achieve private learning with measurable privacy guarantees. This is crucial to privacy as a model that is trained to detect cancer cells when published can inadvertently reveal information about the patient data it was trained on [12]. With recent development on the synergies between privacy and learning, a recent extension of PATE [13] has refined how different, coordinated ML models can be used to simultaneously improve both the accuracy and privacy of the model resulting from the PATE framework.

The extended PATE architecture introduces a new mechanism termed – Confident Aggregator. It utilizes the techniques discussed above and formalizes a selective model. Taking an example of a teacher-student relationship, teachers respond only to the queries made by the students when the consensus among teachers about the is sufficiently high. If the votes assigned to the decision amongst the teachers is larger than a set threshold which is randomized between a range to preserve privacy, we accept the student's query else it's rejected. After the query has been accepted, we proceed with the original noisy aggregation technique by adding noise to each of the vote counts corresponding to each decision and returning the decision with the most votes maintaining a dynamically updating privacy model. As this model filters out the queries that do not require prediction, the Confident Aggregator drastically reduces the privacy budget [13] that is spent using the original technique of single level noisy aggregation.

## II. THE FEDRATED ECOSYSTEM

### A. Understanding Federated Learning

Adoption of AI in industries is facing two primary challenges. First, data exists in fragments. This requires the first step to analyze such data and look at the big picture to be, aggregation. Second, sharing of this data across organizations or even different departments. Federated learning is a framework initially proposed by Google [14]. Federated learning is a distributed machine learning approach where instead of learning after aggregating data, we devise methods to learn from subsets of data and then aggregate the learnings allowing models to operate locally and accurately on mobile devices.

There are $\mathcal{N}$ participants $\{ \mathcal{P}_1, \ldots, \mathcal{P}_\mathcal{N} \}$, who are selected to train the ML model by aggregating their respective data $\{ \mathcal{D}_1, \ldots, \mathcal{D}_\mathcal{N} \}$. The traditional technique suggests aggregating all the data together using $\{ \mathcal{D} = \mathcal{D}_1 \cup \ldots \cup \mathcal{D}_\mathcal{N} \}$ to train a model $\mathcal{M}_{SUM}$. In the FL system the participants collectively train a model $\mathcal{M}_{FED}$ where any participants $\mathcal{P}_i$ does not reveal its data $\mathcal{D}_i$. The accuracy of $\mathcal{M}_{FED}$, denoted as $\delta_{FED}$ should be very close to the performance of $\mathcal{M}_{FED}$, $\delta_{SUM}$. Let $\delta \in +ve\ \mathbb{R}$, then if

$$| \delta_{FED} - \delta_{SUM} | < \delta$$

then the Federated Learning system has a δ-accuracy loss.

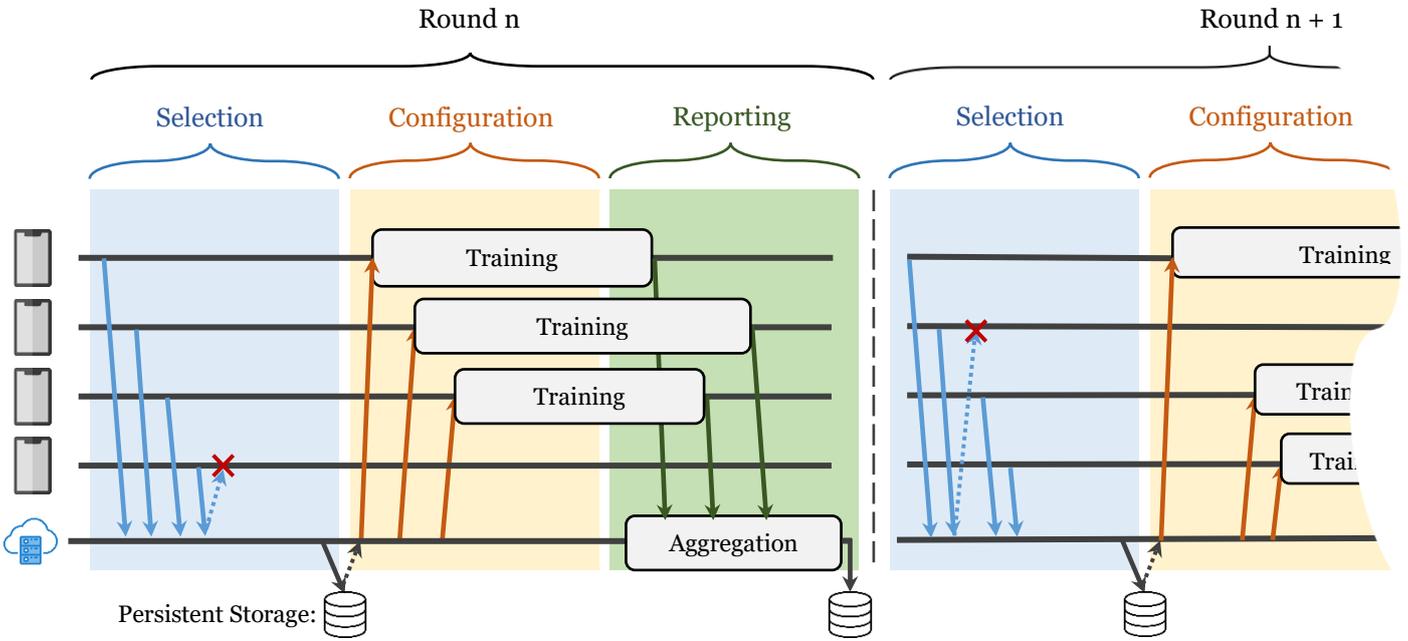

*Figure 1: FL Architecture – Protocol Layer*

## B. The Federated System Design [14]

### 1) Protocol Layer

With reference to *Figure 1*

At the protocol level the interactions happen between a distributed network of mobile devices and the FL Server which is a Cloud based model handling, distribution and aggregation service. The devices connected to the network announce that they are ready to run an FL Task, which is a computation required to train the global Federated Learning Model. It contains important data with respect to the training required to be performed like hyperparameters or evaluation of trained models on local device data. From all the devices that announce availability to the network within a specific period of time, the server selects a subset of these to execute the FL Task. With the round established, the server next sends to each device the current global model parameters and any other necessary details. Each device then performs local computation based on the global model state and its local dataset sending an update in the form of an FL checkpoint back to the server incorporating it in the global model.

### 2) Device Application Layer

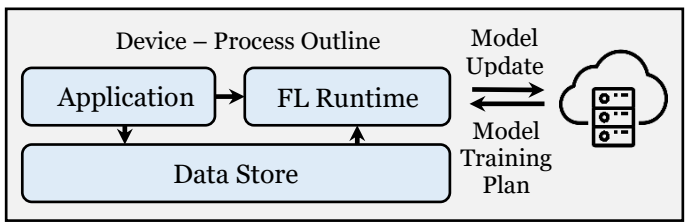

*Figure 2: FL Architecture – Device Layer*

The device maintains an on-device repository of data relevant for data computations and model training. This could be stored in any form ensuring that it follows the guidelines mentioned above to preserve on-device user privacy, for example: limiting the numbers of datapoints stored and cleaning up old data after a set expiration time. This data store should also strictly adhere to platform recommended encryption levels to ensure protection from malware attacks.

Execution Flow – The FL Runtime registers a background worker in a separate process that invokes only when the phone is idle, charging, and on an unmetered network. The worker frees up any allocated resources if any one of these conditions is not met. The worker then connects to the server and requests the metadata for the respective job and runs the required computations. The application utilizes Android's [14] Multi-Tenancy to allow multiple computations to run simultaneously and Attestation to ensure only genuine devices participate in the network with the need to uniquely identify the device to protect against attacks to influence the result from non-genuine devices, protecting against data poisoning.

### 3) Server Architecture

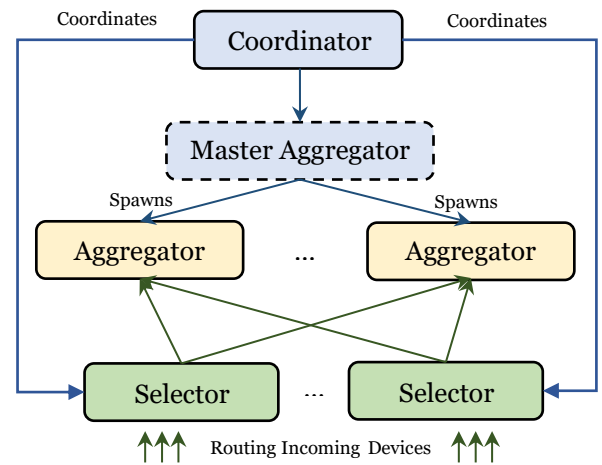

*Figure 3: FL Architecture – Server*

The Federated Learning architecture of the server is based on the Actor Programming Model [15]. Actors, which use message passing as their sole method of communication are the universal primitives enabling concurrent communication and computation. Each actor handles a sequential stream of messages which overall creates a simple model. With multiple instances of these actors running in parallel, the network can handle both vertical and horizontal scaling which will be covered in later sections. The actor can respond by making local decisions, send messages to other actors, or spawning more actors dynamically.

These actors have specific tasks in the system. Coordinators are at top-level maintaining global synchronization. There are multiple Coordinators where each one is responsible for a certain task. The Coordinator registers itself and informs about the devices included in the task to a Selector. The Master Aggregator then spawns Aggregators to manage the rounds of each FL task. Selectors are responsible for accepting and routing connections. After the Master Aggregator and set of Aggregators are spawned, the coordinators act as the load balancers and instruct the Selectors to forward a subset of its connected devices to the Aggregators, allowing the Coordinator to efficiently allocate devices. Master Aggregators handle the scaling by dynamically spawning more aggregators as required by the computation and available devices.

*C. Understanding a Federated Learning Problem.*

To quantify Federated Learning based operations at scale we need to look at the distribution characteristics [16] of the data.

A matrix $\mathcal{D}_n$ represents the data held by each participant n. Each row of the matrix denotes a sample, and each column of the matrix denotes a feature. The dataset might contain label data. The feature $\mathcal{X}$, label $\mathcal{Y}$ and samples $\mathcal{S}$ make up the complete training dataset $(\mathcal{S}, \mathcal{X}, \mathcal{Y})$. The dimensions of the feature and sample space of the dataset might not be identical which helps us classify federated learning into horizontal federated learning, vertical federated learning and federated transfer learning based on the distribution of data amongst various participants.

*1) Horizontal Federated Learning*

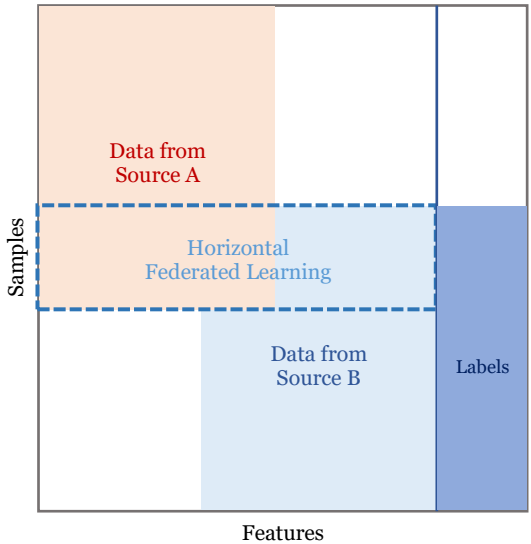

Figure 4: Horizontal Federated Learning

Horizontal federated learning, also known as sample-based [16] federated learning is defined when the datasets have the same feature space but different sample space. For example, two banks may have very different customer base but primarily serve the functions and hence have a common feature base. [17] proposed a collaborative deep learning approach where participants train independently, sharing only a subset of update of parameters. In 2017, Google proposed a horizontal federated learning solution for Android based devices [18] wherein a single user's Android device updates the model parameters locally and uploads the parameters to the Android cloud, thus jointly training the centralized model together with other data owners. Various techniques of secure aggregation scheme discussed above are implemented to protect the privacy of aggregated user updates under their federated learning framework.

A newer framework [19] utilizes *additively homomorphic* encryption for model parameter aggregation to provide security against the central server. Model Parameter Security and encryption are discussed in further sections of the paper. A pipelined federated learning system is proposed to allow multiple aggregators to complete separate tasks, while sharing knowledge with public verifiability using zero knowledge proofs and preserving security. This implementation of concurrent computational learning model can in addition address high communication costs and fault tolerance issues.

Horizontal federated learning can be summarized as:

$$\mathcal{X}_i = \mathcal{X}_j, \mathcal{Y}_i = \mathcal{Y}_j, \mathcal{S}_i \neq \mathcal{S}_j \ \forall \ \mathcal{D}_i, \mathcal{D}_j, i \neq j$$

*2) Vertical Federated Learning*

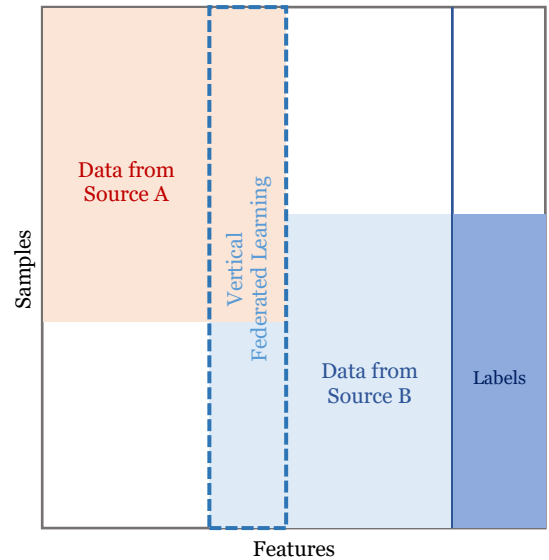

Figure 5: Vertical Federated Learning

Multiple privacy-preserving machine learning algorithms have been well documented for vertically partitioned data, including Cooperative Statistical Analysis [20], association rule mining [21], secure linear regression [22]. A newer model for a vertical federated learning scheme [23] to train a privacy-preserving logistic regression model where the authors studied the effect of entity resolution on the learning performance and applied Taylor approximation to the loss and gradient functions so that homomorphic encryption can be adopted towards privacy-preserving computations in the entire network.

Vertical federated learning or feature-based federated learning (depicted in Figure 5) is implemented where two datasets have the same sample space but different feature space. For example, two different companies are in the same city, one is a bank, and the other is a restaurant. Their userbase is likely to contain most of the residents of the area, so the intersection of their user space is large. However, since the bank records the user's revenue and expenditure behavior and credit rating, and the restaurant retains the user's purchasing history, their feature spaces vary widely.

Vertical federated learning is the process of consolidating different features and computing the training loss and gradients in a privacy-preserving manner building a collaborative model with data from both participants. In such a federal ecosystem, the identity and the status of each participating party is the same. Therefore, in such a system, we have,

$$\mathcal{X}_i \neq \mathcal{X}_j,\ \mathcal{Y}_i \neq \mathcal{Y}_j,\ \mathcal{S}_i = \mathcal{S}_j\ \forall\ \mathcal{D}_i,\ \mathcal{D}_j,\ i \neq j$$

*3) Federated Transfer Learning*

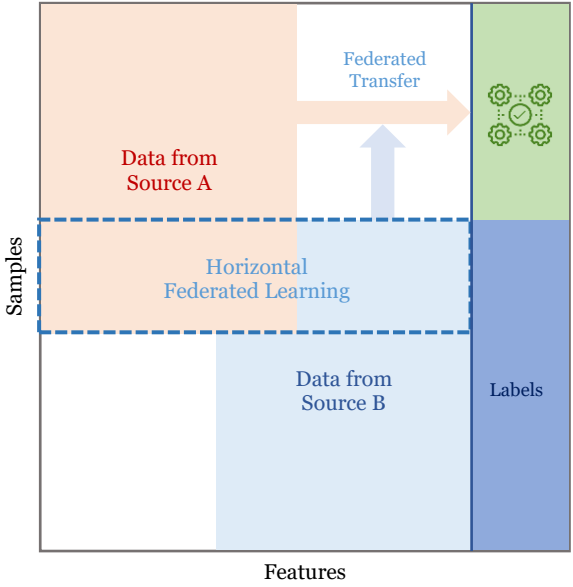

*Figure* 6: Federated Transfer Learning

Federated Transfer Learning can be utilized in scenarios when two datasets have different sample and feature space. Consider two institutions, one is a bank located in India, and the other is a restaurant located in The United States. Considering geographical restrictions, the userbase of the two institutions have a very small intersection. As these are different businesses, only a small portion of the feature space from both parties are found in the intersection. In these cases, transfer learning [24] algorithms can be applied solve for the entire sample and feature space under the federated architecture shown in Figure 6. A common representation between two feature spaces is learned using the limited common datapoints available in the dataset and later applied to solve for predictions for samples. Federated Transfer Learning can be summed up as,

$$\mathcal{X}_i \neq \mathcal{X}_j,\ \mathcal{Y}_i \neq \mathcal{Y}_j,\ \mathcal{S}_i = \mathcal{S}_j\ \forall\ \mathcal{D}_i,\ \mathcal{D}_j,\ i \neq j$$

*D. Maintaining Accuracy and Security at Scale.*

Recent research [18] has proposed building a secure client-server model where the FL system segments the data sent by the users, allowing the models at the client level to collaborate and build a global federated model. Such a federated method of model building ensures no data leakage with further improvements as suggested in [25], propose methods to cut down on the communication cost and facilitate training over mobile clients. Compression technologies also come into play to improve the efficiency of the entire system like Deep Gradient Compression [26] greatly reduces the communication bandwidth threshold in when training a distributed model at scale.

A federated system at scale has the following characteristics,

- *Massively Distributed Data*: Data points are stored across a very large number of participants $\mathcal{N}$. To be specific, the number of participants can be much bigger than the average number of data points by any given participant $\left(\frac{\mathcal{D}}{\mathcal{N}}\right)$.
- *Non-IID (Independent or Identical) Data*: This means that data points available on-device are far from being a proper representative sample of the overall distribution of data leading to a highly skewed non-IID data distribution. This leads to loss in accuracy and degradation of the aggregated FL model.
- *Unbalanced* $\mathcal{D}$: Different participants in the network might contain varying amounts of on-device data available for training.

This accuracy reduction can be explained by the weight divergence, which can be quantified by the earth mover's distance (EMD) [5] or the t-closeness between the distribution of on-device dataset. To further optimize learning from local subsets and preventing the consolidated model from skewing, we utilize Federated – Stochastic Variance Reduced Gradient (SVRG) which is derived from a combination of research on the Stochastic Variance Reduced Gradient (SVRG) [27] [28], a stochastic method with explicit variance reduction, and the Distributed Approximate Newton (DANE) [29] for distributed optimization.

To define Federated – SVRG [25] we consider the following parameters,

- $\mathcal{N}$ − total number of data points utilized for training.
- $\mathcal{P}_k$ − set of indices corresponding to data points stored on any device $\mathcal{K}$.
- $n_k = |\mathcal{P}_k|$ – number of data points stored on a device k.
- $n^j = \left|\{i \in \{1, \ldots, n\}:\ x_i^T e_j \neq 0, j \neq 0\}\right|$ – Number of data points.
- $n^j{}_k = \left|\{i \in \mathcal{P}_k:\ x_i^T e_j \neq 0\}\right|$ – the number of data points stored on node $k$ with nonzero $j^{th}$ coordinate
- $\phi^j = \frac{n^j}{n}$ – frequency of appearance of nonzero elements in $j^{th}$ coordinate.
- $\phi^j{}_k = \frac{n^j{}_k}{n}$ – frequency of appearance of nonzero elements in $j^{th}$ coordinate on node $k$.
- $s^j{}_k = \frac{\phi^j}{\phi^j{}_k}$ – ratio of global and local appearance frequencies on node $k$ in $j^{th}$ coordinate.
- $\omega^j = \left|\{\mathcal{P}_k: n^j{}_k \neq 0\}\right| -$ Number of nodes that contain data point with nonzero $j^{th}$ coordinate.
- $a^j = \frac{\mathcal{K}}{\omega^j} -$ aggregation parameter for coordinate $j$
- $A = Diag(a^j) -$ A diagonal matrix with $a^j$ as $j^{th}$ element on the diagonal.

- $\hbar_k = \frac{\hbar}{n_k}$ – Local step-size.
- $\frac{n_k}{n}(\omega_k - \omega^t)$ – Consolidation of updates proportional to partition sizes.
- $S_k$ – Scaling stochastic gradients by diagonal matrix.
- $S_k$ – Per-coordinate scaling of aggregated updates.

---

**Algorithm 1**: Federated – SVRG

1: **parameters**: $\hbar$ = step-size, data partition $\{\mathcal{P}_k\}_{k=1}^{K}$, diagonal matrices $S_k \in \mathbb{R}^{d \times d}$ for $k \in \{1, \dots, K\}$
2: **for** s = 0, 1, 2, 3, 4, … **do**
3:     Compute $\nabla f(\omega^t) = \frac{1}{n}\sum_{i=1}^{n} \nabla f_i(\omega^t)$
4:     **for** k = 1 *to* K **do in parallel** over nodes k
5:         Initialize: $\omega_k = \omega^t$ and $\hbar_k = \frac{\hbar}{n_k}$
6:         Let $\{i_t\}_{t=1}^{n_k}$ be random permutation of $\mathcal{P}_k$
7:         **for do**
8:            $\omega_k = \omega_k - h_k(S_k[\nabla f_{i_t}(\omega_k) - \nabla f_{i_t}(\omega^t)] + \nabla f(\omega^t))$
9:         **end for**
10:    **end for**
11:    $\omega^t = \omega^t + A\sum_{k=1}^{K} \frac{n_k}{n}(\omega_k - \omega^t)$
12: **end for**

---

Such an optimized Federated Learning System can be combined with cryptographic approaches to secure user data and create an end to end encrypted system that ensures privacy and security.

Homomorphic Encryption plays an important part in the security of this two-stage communication channel by adding a third stage of *additively homomorphic* [23] encryption. This secures the network against an honest-but-curious adversary, allowing to learn without either exposing user data or sharing which identifier the data providers have in common. *Additively homomorphic* encryption maintains these standards at scale to solve computational problems, with millions of datapoints [23] each with hundreds of features.

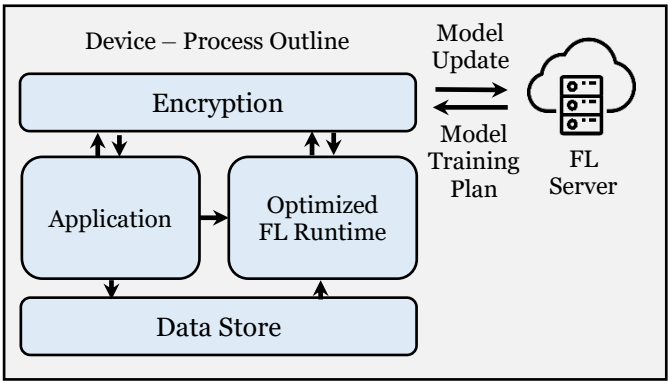

*Figure 7: Encrypted & Optimized FL System Device Layer*

The benefits of an additively homomorphic encryption scheme are that an operation that produces the encryption of the sum of two numbers, given only the encryptions of the numbers. For example,

Let the encryption of a number $x$ be $\|x\|$. For simplicity we overload the notation and denote the operator with '+' as well. For any unencrypted $x$ and $y$ we have,

$$\|x\| + \|y\| = \|x + y\|$$

Therefore, we can also multiply an encrypted and an unencrypted together by repeated addition,

$$y \times \|x\| = \|yx\|$$

where $y$ is unencrypted. Hence, we can calculate the sum and product of any unencrypted and encrypted datapoint without leaving the domain of encryption. These operations when extended to work with matrices and vectors, for example, denoting the inner product of two vectors of unencrypted values $x$ and $y$ by $y > \|x\| = \|y > x\|$ and the individual products like $y \circ \|x\| = \|y \circ x\|$. Summation and matrix operations work in a similar manner.

Therefore, an *additively homomorphic* encryption scheme can be utilized to implement useful linear algebra primitives for various machine learning operations. This mostly removes the need to operate on unencrypted values ensuring user privacy and maintaining the security of the network by making sure that no identifier leaves the device unencrypted. These tasks may be computationally intensive and can be further optimized by utilizing privacy-preserving entity resolution [23].

### III. BLOCKCHAIN INTERGRATION

Blockchain technology enables decentralized data sharing in a transactional manner across a large network of untrusted participants. It allows us to create a new form of distributed software architecture, where individual components can find agreements on their shared states without trusting a central integration point or any individual participant in the network. Blockchain technology can be utilized as a software connector [30] leading to important architectural considerations on the resulting performance and on important network attributes like security, privacy, scalability and sustainability.

Utilizing blockchain technology behind a privacy preserving network could improve data transparency and traceability. Combining such a privacy focused federated learning ecosystem with a blockchained architecture where local models can be exchanged and verified would enable the federated learning server to exist in a decentralized manner without the need for a centralized coordinator by utilizing the blockchain's own consensus mechanism.

A Centralized FL operation has primarily, two issues.
1. It depends upon a single centralized server to coordinate the entire FL ecosystem. This makes the entire ecosystem vulnerable to server malfunction and bias. This could possibly result in an inaccurate global model with updates distorting all local model too.
2. Local devices have no incentive to contribute more than the global average number of sample data points contributed leading to devices with a larger contribution to the network being put at the same level as the devices with a lot less contribution.

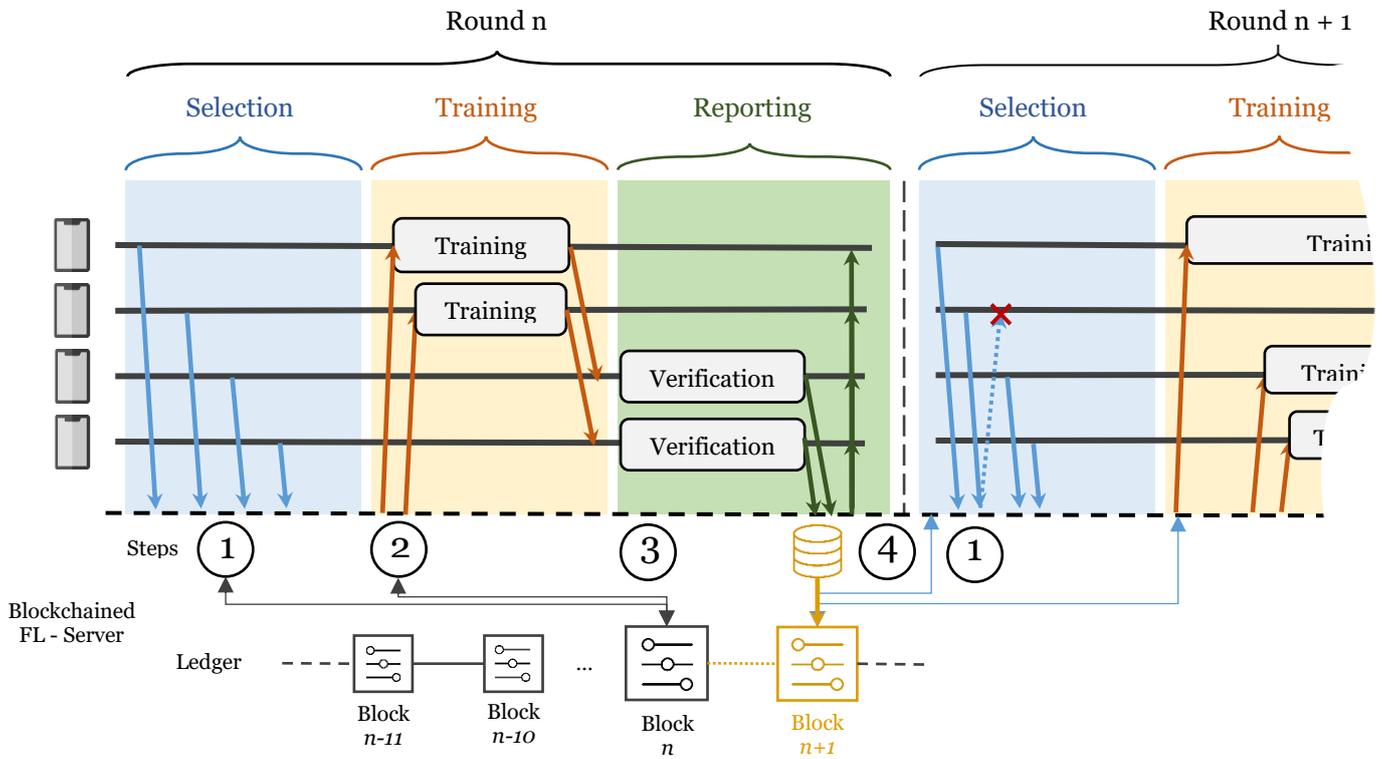

*Figure 8: Blockchain backed FL Architecture*

These problems can be solved by leveraging the distributed architecture of a blockchain instead of a centralized coordination server. Such a blockchain network enables exchanging devices' local model updates while verifying their work with the need of a central server. This architecture overcomes the problem if a single point of failure. This also enables untrusted devices in a public network to participate, as all results pass through a layer of validation. The creation of blocks allows the participants to be rewarded for their contribution to the network proportional to their training sample sizes or the computational power provided to the network depending upon the tasks assigned to the devices by the FL Plan residing on the blockchain. A blockchain backed federated learning network promotes the federation of more devices with a larger number of training samples and devices which have extra computational abilities.

This Federated Learning process has been broken down into the following steps,

*1) Design of the FL Plan*

The parameters of the model to be trained, the specification of devices required by the network, the metadata required to execute the instructions, the privacy parameters and the composition of the entire model gets stored as a FL plan on the blockchain, each containing the relevant training parameters like the privacy budget public. The FL Plan acts the rule book, by which the entire training process and the devices that participate in the training abide by. Each FL plan consists of FL Tasks that are required to be computed to train the global FL model.

*2) Device Categorization*

The parameters stored in the FL Plan together with the participant decide the kind of task that should be assigned to the device (depicted in Step 1 of Figure 8). After which the FL Task is sent to the respective device to compute. At the same time a device is associated with the task to verify the computations known as a miner.

*3) Task Execution*

As depicted in Step 2 of Figure 8, the FL Plan is executed by a set of devices participating in the learning round each with a specific FL Task. Each device has a set of data samples and trains its local model with the objective of minimizing the loss function f(w) for a global weight vector w. The example of these loss functions can be seen in [31]. Utilizing the Federated–SVRG algorithm specified above (Algorithm 1) with additive homomorphic encryption the model is locally trained. The updates from this trained local model contained on the device are then sent to its associated miner that was randomly selected out of a set of miners. The network solves the entire FL plan parallelly across all devices associated to the network. The updates are then aggregated using the Distributed Approximate Newton [29] method.

*4) Result Verification*

The verification of results is then computed by exchanging the local model updates truthfully through the distributed ledger. Each block in the ledger (Figure 8) consists of the local model updates of the devices at the $n^{th}$ round (shown in, Step 3 of Figure 8) and its local computation time along with the FL Plan. The miners broadcast the obtained local model updates. Each miner has a candidate block that is filled with the verified local model updates from its associated devices and/or other miners. This filling procedure continues until it reaches the block size. Following the PoW [32] approach, the miner randomly generates a hash value by changing its input number, i.e., nonce, until the generated hash value becomes smaller than a target value. Once the miner succeeds in finding the hash value, its candidate block can be the new block. This PoW consensus mechanism can also be changed to another consensus mechanism like Proof-of-Stake to reduce the overall computational overhead required to execute the FL Plan. The generated block is then propagated to all other miners. The block propagation delay can be minimized by modifying the block generation rate according to the delay.

The blockchain network also provides rewards to the devices and for the verification process to the miners. The rewards are distributed according to the task executed by the devices. The miner gets the reward by mining the block and the devices which contributed the data samples get a reward proportional to the number and quality of samples they contributed to the network. As the global model update is computed locally on-device the entire distributed FL system is secured against malfunction of any single device and prevents excessive computational overheads for miners participating in the network.

### B. Privacy-Preserving Data Sharing for Consortia

When optimizing the Blockchained FL Network for consortia, we need to take into account some important changes. One of the foremost being in the data distribution. A consortium will contain less participants as compared to a public network, but each of these participants will contain much larger on-device samples. Second, these participants will have high power server-grade equipment connecting to the network instead of portable mobile devices. These devices will almost always satisfy the conditions that the device must be connected to an un-metered connection and be charging. Due to large data sizes a distributed storage infrastructure like IPFS will have to be integrated to ensure no single entity has access to the entire encrypted data. When working with consortia, it is critical to abide by important legal and industrial requirements.

Implementing this network technology can be complex and infrastructure intensive for small scale organizations and startups. Therefore, it can be a good choice to utilize open source distributed technologies like Ethereum and use their existing decentralized infrastructure coupled with IPFS, or Hyperledger Besu – which is also based on Ethereum to build enterprise-ready secure architecture from the ground up.

### C. Adapting the FL Ecosytem to Existing Technologies

Ethereum [33] which is powered by a P2P consensus protocol, becomes ideal for building a completely decentralized FL system. Ethereum supports a system of Smart Contracts which act as the rulebook that Decentralized apps (dApps) build on top of the Ethereum platform provide a robust platform. FL Plans reside inside these Smart Contracts which as facilitators. This also allows the aggregation step to be done autonomously by the devices by utilizing the global copy of the model from the chain and updating it independently without the need for direct dependence on a miner to push the verified model parameters.

Layer-2 Scaling [34] proposes a technique using permissionless side chains with merged block production similar to merged mining in a trustless manner. This allows multiple FL plans to be executed parallelly across multiple sidechains, drastically reducing block generation times and gas costs. Projects like Matic Network allow up to 65 thousand transactions per second [35] increasing the throughput of blockchain dependent network decisions drastically.

## IV. FEDERATED LEARNING WITH IOT DEVICES

The Internet of Things (IoT) is leading a paradigm shift towards adding more intelligence and connectivity to the objects that surround us. Everything from smart fridges to thermostats can connect to the internet, generating massive amounts of data that can become a real time support towards deriving newer inferences and better prediction models providing newer and innovative techniques to create meaningful experiences.

However, IoT devices come with certain issues. Most devices have massive constraints like computing power constraints, storage constraints and low battery power. Heterogeneity of IoT systems in terms of devices, communication protocols, data types also become the root cause of other challenges such as interoperability amongst IoT devices. Privacy and security vulnerabilities are also a problem. Blockchain technology can offer a potential solution to challenges like poor interoperability, privacy and security vulnerabilities and is also improve heterogeneity across IoT systems.

An IoT device can be utilized as a real time data stream to provide crucial datapoints, for example – real-time metrics across a city on temperature and humidity could be used to train a global federated learning model to predict the weather. The metrics generated by IoT devices are encrypted using onboard additively homomorphic encryption and sent to a device participating in the network for further computations. Encryption and on-device differential privacy with model parameters available in the FL plan on the Blockchain. ensures that the data cannot be linked back to the device. The device then utilizes this new data and trains the local FL model. After training the updated model gets written on the blockchain. This updated global model can then even be utilized by the IoT devices to predict local weather conditions and alert the people residing locally of the upcoming weather in a privacy-preserving, secure and trustless manner.

Security can be further improved in IoT devices by utilizing Threshold Secret Sharing (TSS) [36] to segment the information into pieces and distribute them amongst multiple devices in the network so that the information can only be retrieved collaboratively by groups of devices. This ensures the privacy and integrity of the data, even if attackers hijack a large number of devices.

## V. SUMMARY

In this work we discuss the various segments of a blockchain based privacy-preserving federated learning network. This network allows us to learn and share sensitive user data in a private, federated and encrypted manner to train a globally coordinated FL model. This distributed learning architecture is backed by a blockchain network that removes the need for a centralized server, therefore removing any single point of failure problems. With newer advances in blockchain technology allowing the network to scale, allows the network to handle thousands of devices simultaneously. This paper also shows how this network can be incentivized to allow devices to be rewarded for their work, allowing devices ranging from low powered IoT Devices, to massive server computer networks to participate in the global network. With support for integration with Ethereum, this allows any organization to build robust privacy-preserving networks without worrying about maintaining a blockchain backbone.